\documentclass[12pt,preprint]{aastex}

\begin{document}
\title{New Completeness Methods for Estimating Exoplanet Discoveries by Direct Detection}

\author{Robert A.\ Brown and R\'emi Soummer}

\affil{Space Telescope Science Institute, 3700 San Martin Drive, Baltimore, MD 21218}

\begin{abstract}
We report new methods for evaluating realistic observing programs that search stars for planets by direct imaging, where observations are selected from an optimized star list, and where stars can be observed multiple times. We show how these methods bring critical insight into the design of the mission and its instruments. These methods provide an estimate of the outcome of the observing program: the probability distribution of discoveries (detection and/or characterization), and an estimate of the occurrence rate of planets ($\eta$). We show that these parameters can be accurately estimated from a single mission simulation, without the need for a complete Monte Carlo mission simulation, and we prove the accuracy of this new approach. Our methods provide the tools to define a mission for a particular science goal, for example defined by the expected number of discoveries and its confidence level. We detail how an optimized star list can be built and how successive observations can be selected. Our approach also provides other critical mission attributes, such as the number of stars expected to be searched, and the probability of zero discoveries. Because these attributes depend strongly on the mission scale (telescope diameter, observing capabilities and constraints, mission lifetime, etc.), our methods are directly applicable to the design of such future missions and provide guidance to the mission and instrument design based on scientific performance. We illustrate our new methods with practical calculations and exploratory design reference missions (DRMs) for the \textit{James Webb Space Telescope} (\textit{JWST}) operating with a distant starshade to reduce scattered and diffracted starlight on the focal plane. We estimate that 5 habitable Earth-mass planets would be discovered and characterized with spectroscopy, with a probability of zero discoveries of 0.004, assuming a small fraction of \textit{JWST} observing time (7\%), $\eta=0.3$, and 70 observing visits, limited by starshade fuel.

\end{abstract}

\section{Introduction}\label{sec1}

Various proposals for direct-imaging searches for Earth-like planets are now on the table \citep{Guyon,Cash08,Spergel,Soummerb}. This paper presents methods to guide the development of the new instrumentation that such searches will require. They can also help optimize observing programs by scheduling observations to maximize their impact. Using these new methods, we can predict the outcome of a single observation or an entire observing program, and interpret the observational results. These tools should be widely useful for studying, comparing, and optimizing alternative direct-search concepts.

The completeness of a  direct limiting search observation (LSO) is the fraction of all possible planets of interest (POIs) that satisfy the detection criteria. By definition, an LSO has sufficient exposure time, $t_{\mathrm{exp},j}$ on the $j^{\mathrm{th}}$ star, to reach the systematic limit of the instrument. For optical missions, an LSO is assumed to achieve the desired photometric signal-to-noise ratio on the limiting source, which has magnitude \textit{mag}$_{j}+\Delta\textit{mag}_{0}$, where \textit{mag}$_{j}$ is the stellar magnitude, and $\Delta\textit{mag}_{0}$ is the limiting magnitude difference with the star (flux contrast), expected to be determined by speckle instability \citep{Brown05}.

In the simplest picture, an LSO discovers any and all POIs that satisfy two criteria at the time of the observation:
\begin{equation}
\label{eq1}
\Delta\textit{mag}<\Delta\textit{mag}_0~,
\end{equation}
and 
\begin{equation}
\label{eq2}
s>\textit{IWA}~,
\end{equation}
where $s$ is the angular separation between planet and star, and \textit{IWA} is the inner working angle (angular radius of the real or effective central field obscuration). This simplified picture of a sharp dividing line between detectable and undetectable planets has proven useful for measuring search power to first order. Fidelity could be improved, if necessary, by adding more complex detection criteria, such as a detection probability that varies in a more complicated and realistic way over the field of view. 

In a common treatment, which we follow here, the POIs are body-twins of Earth with orbits in the habitable zone. 

Let a fraction $\eta$ of all stars in the universe have a POI. The list of candidate stars for the observing program is a subset of all stars, and within this subset, the expected fraction is $\eta$, but the actual fraction will vary. Also, we have limited knowledge of $\eta$, a number that is both interesting scientifically and useful operationally to schedule observations to maximize discoveries. Therefore, we will draw distinctions when necessary between the true value ($\eta_\mathrm{true}$), which is unknown, the assumed value for science operations ($\eta_\mathrm{ops}$), which is used to estimate discovery probabilities, and the value estimated from the results of a search program [$E(\eta)$].

Our new methods extend the original concept of direct-search completeness for exoplanets \citep{Brown04,Brown05} for purposes of optimizing the timing of revisits to previously searched stars, increasing the realism of instrument comparisons, and providing an estimate of $\eta$ from the results of observing programs.  Meanwhile, \citet{Brown09a,Brown09b} has extended completeness studies to indirect detection by reflex astrometry, photometric detection in the case of no occultations, and the estimation of orbital parameters from Keplerian data sets.

\section{Evolution and Estimation of Completeness}\label{new2}

\subsection{Four Types of Completeness}\label{sec2}

Four types of completeness pertain to a program of LSOs: virgin, dynamic, accumulated, and ultimate. 

Until now, the scheduling of LSOs in mission studies of direct-search power has been based on ``virgin,'' time-independent, first-visit completeness $c_{1,j}$, which is the completeness of the first LSO of star~$j$ \citep{Brown04,Agol07,SKC10}. In this study, we include the possibility of multiple LSOs of any target star.

A non-detection by an LSO rules out some fraction of possible planets: those with sufficient angular separation and brightness at the time $t$ of the LSO to be detectable---if they existed. After the LSO, a pool of possible planets may remain, comprising planets that had not been ruled out by previous LSOs and also were not detectable by the most recent LSO. As time goes on, each planet in this pool moves along its unique orbit and may become detectable at some future time. In this way, the fraction of all possible planets that is detectable on the $i^\mathrm{th}$ visit to the $j^\mathrm{th}$ star at time $t$---which is the dynamic completeness, $c_{i,j} (t)$---depends on the elapsed time since each of the $i-1$ preceding LSOs. 

Accumulated completeness ($C_{i,j}$) is the sum-total completeness of $i$ LSOs of star $j$:
\begin{equation}
\label{eq3}
C_{i,j}\equiv\sum\limits_{l=1}^{i} {c_{l,j} (t_l)}~.
\end{equation}
$C_{i,j}$ increases monotonically with $i$.

Ultimate completeness ($C_{\infty,j})$ is the maximum value of $C_{i,j}$. It is the value of completeness that would be accumulated from an arbitrarily high number of LSOs spread arbitrarily over time: 
\begin{equation}
\label{eq4}
C_{\infty,j} =\sum\limits_{l=1}^\infty {c_{l,j} (t_l)}~. 
\end{equation}
$C_{\infty,j}<1$ whenever some POI orbits are permanently fainter than $\Delta\textit{mag}_{0}$ or permanently obscured inside \textit{IWA}---or never brighter than $\Delta\textit{mag}_0$ and located outside \textit{IWA} at the same time.

\subsection{Estimating Dynamic Completeness \boldmath{$c_{i,j}$}}\label{sec3}

All types of completeness are derived from $c_{i,j}$, and all results (probability of discovery and mission outcomes) are ultimately based on this quantity. 

Because the planetary position is determined by a transcendental equation (Kepler's Equation), $c_{i,j}$ must be estimated by Monte Carlo trials. In these trials, we represent the universe of POIs by a large random sample of $N_{0}$ \textit{particular} POIs, each of which is defined by randomly chosen values for ten parameters: $\{a$, $e$, $M_{0}$, $i$, $\omega_\mathrm{p}$, $\Omega$, ${\cal T}$, $R_\mathrm{p}$, $q$, $\Phi\}$, where $a$ is the semimajor axis, $e$ is the orbital eccentricity, $M_{0}$ is the mean anomaly at some definite time, $i$ is the inclination angle, $\omega_\mathrm{p}$ is the argument of periastron, $\Omega$ is the position angle of the ascending node, ${\cal T}$ is the orbital period, $R_\mathrm{p}$ is the planetary radius, $q$ is the geometric albedo, and $\Phi$ comprises the necessary sub-parameters for defining the phase function. 

In this paper, the POIs are Earth-twins on habitable-zone orbits. The particular values of six parameters are drawn from random deviates: $0.7\sqrt{L}\le a\le1.5\sqrt{L}$, $0\le e \le0.35$, and $0\le M_{0} \le 2\pi$ (uniformly distributed; $L$ is the stellar luminosity); $i$, $\omega_\mathrm{p}$, and $\Omega$ uniformly distributed on the sphere. Three parameters are delta functions: $R_\mathrm{p}=R_{\oplus }$, $q$ depends on the filter passband, and $\Phi$ is the Lambertian phase function. ${\cal T}$ is given by $a$ and the stellar mass via Kepler's Third Law.

When estimating $c_{1,j}$, for the \textit{first} LSO, $\Delta$\textit{mag} and $s$ are computed for all $N_{0}$ POIs at, say, $t=0$, when the orbital phases of all POIs in the sample are equal to $\{M_{0}\}$. Thereafter, record is kept of the epoch of each LSO, $t_{i>1}$, only for the POIs that have not yet been eliminated and are still in play. 

For the $i^{\mathrm{th}}$ LSO, we identify and count the number of POIs that satisfy Eqs.~(1--2), $N_{i,j}$, producing: 
\begin{equation}
\label{eq5}
c_{i,j} =\frac{N_{i,j} }{N_0 }~.
\end{equation}
Based on Eq.~(\ref{eq5}), all types of completeness can be computed with any required precision by appropriately choosing the value of $N_{0}$.

\subsection{Rebound of \boldmath{$c_{i,j}$}}\label{sec4}

$c_{i+1,j}(t)$ rebounds following the $i^{\mathrm{th}}$ LSO. $c_{i+1,j}(t_i+\epsilon)=0$, where $\epsilon$ is a diminishingly small increment of time, and then it rebounds  towards a constant, asymptotic value, $c_{i+1,j}(\infty)$, as the orbits of still-possible POIs lose orbital phase coherence. 

\begin{figure}
\plotone{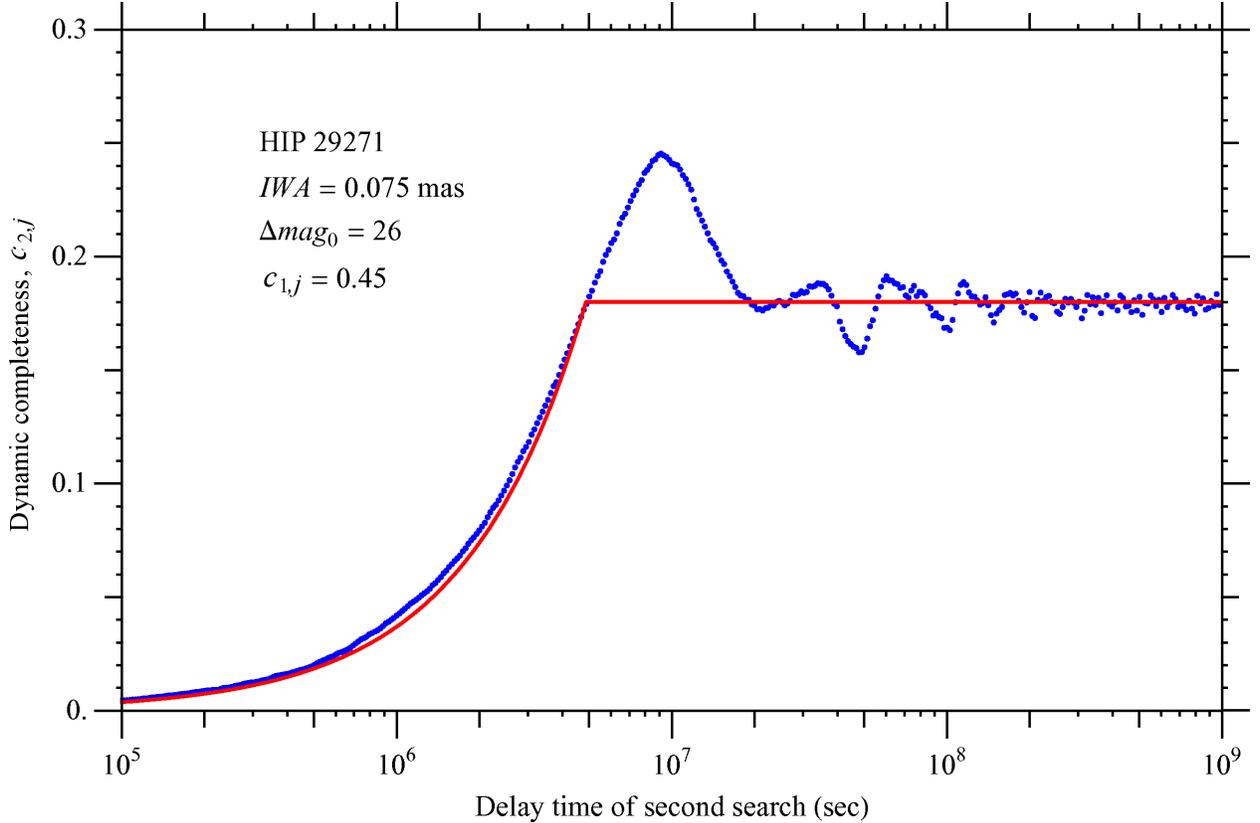}
\label{fig1}
\caption{Typical rebound of dynamic completeness $c_{2,j}(t)$ following the first limiting search observation (LSO) at $t=0$. Blue: values calculated from Eq.~5 with $N_{0}=20,000$, for Earth-like, habitable-zone planets around HIP~29271, assuming \textit{IWA}~= 0.075 arcsec, $q=0.26$, and $\Delta\textit{mag}_0 = 26$. Red: the linear approximation. Here, $c_{2,j}(\infty) = 0.18$ and \textit{breaktime}~$=5\times10^7$~sec.}
\end{figure}

The blue points in Figure~1 show how $c_{2,j}$ rebounds after a first LSO at $t=0$ of HIP~29271 for \textit{IWA}~= 0.075 arcsec \citep[\emph{THEIA;}][]{SKC10} $q=0.26$, and $\Delta\textit{mag}_{0}=26$ \citep[typical value;][]{Brown05}. The rebound resembles the response of the output voltage for an underdamped, series, \textit{LRC} circuit after a step change in the input voltage. After a linear rise (curved on a logarithmic plot), $c_{2,j}$ undergoes damped oscillations. The details---rise time, damping time, and asymptotic value $c_{2,j}(\infty)$---depend on the star's particular mix of habitable-zone orbits resolved by \textit{IWA}. For different stellar mass, luminosity, and distance, for other values of \textit{IWA} and $\Delta\textit{mag}_{0}$, and for other definitions of POIs, the rebound of $c_{2,j}$ will be qualitatively similar to Figure~1, but with different details and numbers.
For example, other factors being equal, a star at larger distance would have a longer rebound time, because the central field obscuration (\textit{IWA}) would limit the observation to planets with wider separations, and therefore a longer time would be necessary to loose orbit coherence.

When the next observation is being planned, we need to know $c_{i,j}(t)$ for each star in play, and where $i-1$ observations of star~$j$ have already been performed. When $i=1$, we would use tabulated values of time-independent virgin completeness, and no real-time computation would be required. When $i>1$, however, we may need an efficient function or procedure for computing the values of $c_{i,j}(t)$. 

The most accurate, brute-force method would perform a blue-point-type calculation (see Figure~1) for every star in play every time a new observation is planned. The number of times would be of order the number of stars times the number of observations. For example, the number of blue-point-type calculations would exceed $10^5$ for a program of 100 stars and 1,000 LSOs, typical for a 4-m class instrument with \textit{IWA}\,$=0.075$ arcsec. Monte Carlo full-mission studies would be impractical, as each of the 400 blue points in Figure~1 took $\sim$5~sec to compute on a 3~GHz Intel Xenon processor running \textsc{Mathematica}~6. Therefore, we must look at two approximate functions for $c_{i,j}(t)$, one of which may be perfectly adequate for first-order scheduling studies. They demand only one or four blue-point-type calculations performed a number of times that is of order the number of observations.

The first alternative approximation is a linear function, $c_{i,j}(t) = \textit{slope} \times t$ for $t<\textit{breaktime}$, and $c_{i,j}(t)=c_{i,j}(\infty)$ for 
$t>\textit{breaktime}$. This is illustrated by the red curve in Figure~1.

We used the following algorithm to find the parameters \textit{slope} and 
$\textit{breaktime}\equiv c_{i,j}(\infty)/$\textit{slope}. The algorithm comprises four blue-point-type computations of $c_{i,j}(t)$. First, we estimate $c_{i,j}(\infty)$ by computing $c_{i,j}(10^{10}$~sec). Second, we compute $c_{i,j}(10^{5.5}$~sec), and use it to make a first estimate of the \textit{slope}, $\textit{slope}1\equiv c_{i,j}(10^{5.5}~\mathrm{sec})/10^{5.5}$~sec. (The starting point $t=10^{5.5}$ is somewhat arbitrary.  It should be large enough to afford an accurate value of $c_{i,j}$ according the counting statistics of Eq.~(5), but also confidently smaller than the true value of \textit{breaktime}.) Third, we compute a first estimate of \textit{breaktime}, $\textit{breaktime}1=c_{i,j}(10^{10}~\mathrm{sec})/\textit{slope}$1, and compute $c_{i,j}(\textit{breaktime}1)$, which produce second estimates of the breaktime and slope:
\begin{equation}
\textit{breaktime}2\equiv\textit{breaktime}1 - \frac{c_{i,j}(\textit{breaktime}1)-c_{i,j}(10^{5.5}~\mathrm{sec})}{\textit{slope}1}~,
\end{equation}
and $\textit{slope}2\equiv c_{i,j}(10^{10})/\textit{breaktime}2$. Fourth, we compute $c_{i,j}(\textit{breaktime}2$)---the fourth and last blue-point-type computation---and use it to produce the final estimates
\begin{equation}
\textit{breaktimeFinal}\equiv\textit{breaktime}2 - \frac{c_{i,j}(\textit{breaktime}2)-c_{i,j}(\textit{breaktime}1)}{\textit{slope}2}~,
\end{equation}
and $\textit{slopeFinal}\equiv c_{i,j}(10^{10}~\mathrm{sec})/\textit{breaktimeFinal}$. (The linear function required about $\sim$20~sec to calculate on a 3~GHz Intel Xenon processor running \textsc{Mathmatica}~6.)

The second alternative approximation is a step function: $c_{i,j}(t) = 0$ for $t<\textit{breaktime}$, and $c_{i,j}(t) = c_{i,j}(\infty)$ for $t>\textit{breaktime}$, where $\textit{breaktime}\sim10^7$~sec for habitable-zone orbits and popular instrument concepts. For first-order investigations, accurate knowledge of the completeness rebound may not be important to the outcome. The important thing is to avoid the mistake that \textit{breaktime} is zero or too small, which error may cause bogus observations to pile up on the high-completeness, low-exposure-time stars. (The step function required $\sim$5~sec to calculate on a 3~GHz Intel Xenon processor running \textsc{Mathmatica}~6.) 

In Figure~2, the first (linear) approximation of $c_{i,j}(t)$ is used to illustrate a hypothetical program of five LSOs to HIP~29271 starting at absolute time $t=10^{6}$~sec. (The abscissa here is now linear.) Such approximations would be used only in the scheduling process of a DRM, where it may be necessary to compute dynamic completeness for many stars on the fly. After the decision observe a particular star, Eq.~(5) would be used to produce an accurate value for the record. Note that as the number of LSOs of a star increases, the accumulate completeness converges on the ultimate completeness, $C_{\infty,j}$.

\begin{figure}
\plotone{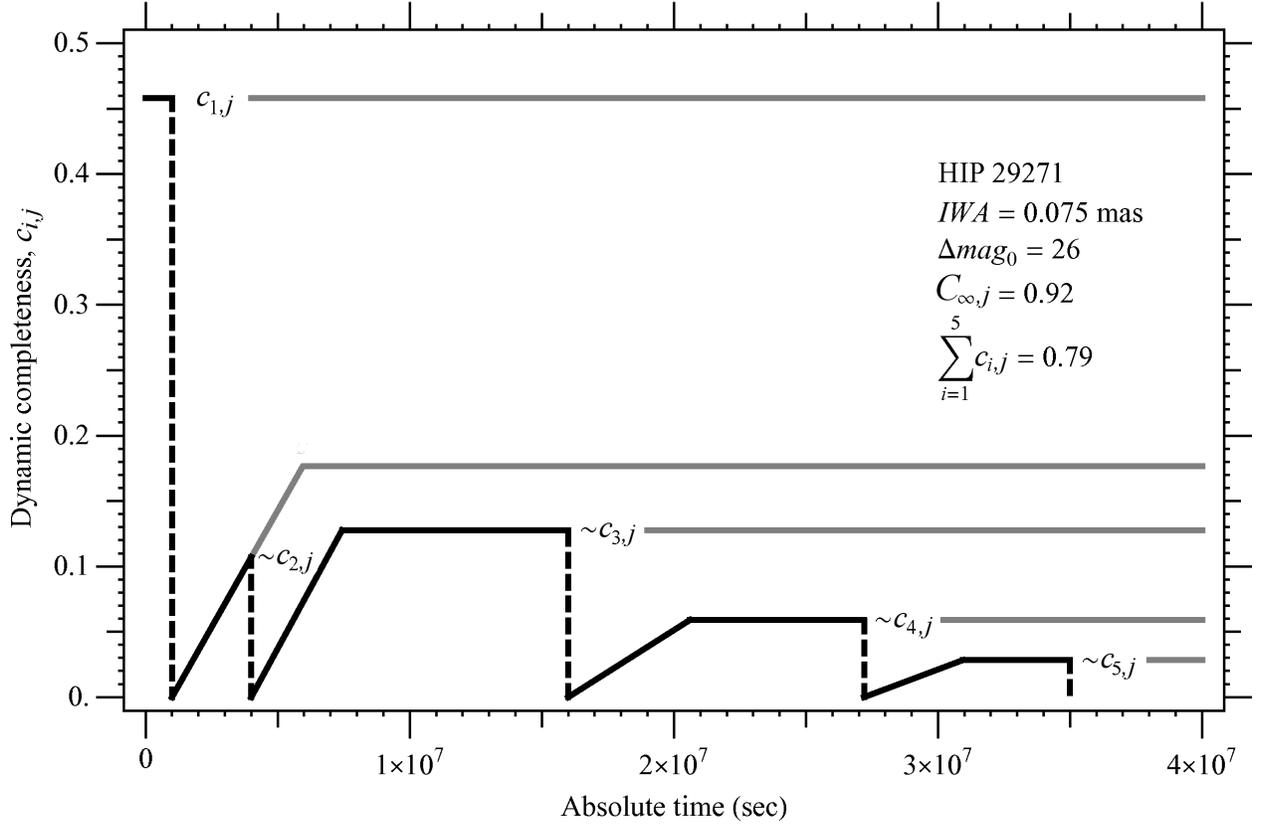}
\label{fig2}
\caption{The dynamic completenesses of five hypothetical LSOs of HIP~29271 at $t=10^{6}$, $4\times10^{6}$, $1.6\times10^{7}$, $2.7\times10^{7}$, and $3.5\times 10^{7}$~sec. The virgin completeness $c_{1,j}=0.46$ is an accurate value, independent of time and computed in advance of simulations, using Eq.~5. The symbols $c_{2-5,j}$ are preceded by tildes to indicate they are approximations. Such approximations would be used only in the selection process of a DRM, where it may be necessary to compute dynamic completeness for many stars on the fly. After the decision to observe a particular star, Eq.~5 would be used to produce an accurate value for the record. Note that as the number of LSOs of a star increases, the accumulate completeness converges on the ultimate completeness, $C_{\infty,j}$.}
\end{figure}

\section{The Probability of a Discovery by the Next LSO, \boldmath{${\cal P}$}}\label{sec5}

After $i-1$ unproductive LSOs of star $j$, ($1 - C_{i-1,j}$) is the fraction of all possible POIs that have not been ruled out and are still in play. The fraction of the \textit{remaining} POIs that the next ($i^{\mathrm{th}}$) LSO at time $t$ would detect is 
\begin{equation}
\label{eq8}
{\cal K}_{i,j} (t)=\frac{c_{i,j} (t)}{(1-C_{i-1,j} )}~.
\end{equation}
The probability of a discovery on the next LSO of star $j$ is
\begin{equation}
\label{eq9}
{\cal P}_{i,j} ={\eta}'_j {\cal K}_{i,j} (t)~, 
\end{equation}
where ${\eta}'_j $ is the Bayesian correction of the occurrence probability $\eta$. ${\eta}'_j $ is the probability that star $j$ possesses a POI after taking into account the contrary evidence of $n_{j}$ previous unproductive LSOs with accumulated completeness $C_{i-1,j}$. 

Bayes's theorem states:
\begin{equation}
\label{eq10}
{\eta}'_j \equiv P({\cal H}\vert {\cal E})=\frac{P({\cal E}\vert {\cal 
H})P({\cal H})}{P({\cal E})}=\frac{(1-C_{i-1,j} )\eta }{1-\eta C_{i-1,j} }~,
\end{equation}
where the hypothesis ${\cal H}$ is that star $j$ has a POI; the evidence ${\cal E}$ is the lack of a discovery so far; $P({\cal E}\vert {\cal H})=(1 - C_{i-1,j})$ is the conditional probability of ${\cal E}$ if ${\cal H}$ is true; $P({\cal H})=\eta $ is the prior probability of ${\cal H}$; and the marginal probability of ${\cal E}$ is 
\begin{eqnarray}
\label{eq11}
P({\cal E})&=&P({\cal E}\vert {\cal H})P({\cal H})+P({\cal E}\vert 
\overline{{\cal H}})P(\overline{{\cal H}})\nonumber\\ 
&=&(1-C_{i-1,j} )\eta +(1-\eta)=1-\eta C_{i-1,j}~,
\end{eqnarray}
where $\overline{\cal H}$ is the hypothesis that star $j$ does not have a POI. [$P({\cal E}\vert \overline{{\cal H}})=1$ and $P(\overline{{\cal H}})=1-\eta$.]

The result for ${\cal P}_{i,j}$ is
\begin{equation}
\label{eq12}
{\cal P}_{i,j} =\frac{\eta c_{i,j} (t)}{1-\eta C_{i-1,j} }~.
\end{equation}
For the case of no prior searches ($i=1,\ C_{0,j}\equiv0$), the probability of a discovery on the first search is $\eta c_{1,j}$, as expected.

We want to confirm numerically that Eq.~12 accurately estimates the probability of a discovery by the next LSO for random values of the parameters, for  example, $\eta = 0.272673$, $C_{1,j} = 0.437671$, and $c_{2,j} = 0.506385$, for which ${\cal P}_{2,j} = 0.156789$. For this verification, we conduct $N_\mathrm{obs} = 200,000$ independent LSOs, each involving $N_0 = 200,000$ possible POIs. Each LSO involves the following computational steps: 
\begin{enumerate}
  \item Randomly pick the serial number of the ``real'' POI: $n = {\cal B}(\eta) {\cal I}(1, N_0)$, where ${\cal B}$ is a Bernoulli random deviate with probability $\eta$, which yields the value 0 or 1, and ${\cal I}$ is a uniform random deviate producing an integer in the range 1 to $N_0$. (If the serial number is zero, it means the ``star'' being observed does not have a POI.)
  \item Perform a first LSO by selecting $N_1 = \textsc{Round}(C_{1,j} N_0)$ random integers in the range 1 to $N_0$, where the \textsc{Round} function yields the closest integer to the argument. If $n$ is one of these $N_1$ integers, then go back and repeat Steps~1 and~2, because we only want cases where the first LSO does not make a ``discovery.'' 
  \item Perform a second LSO by randomly selecting $N_2 = \textsc{Round}(c_{2,j} N_0)$ integers from the $N_0$--$N_1$ integers defined by excluding the $N_1$ integers in Step~2 from the set of all integers 1 to $N_0$. If $n$ is equal to one of these $N_2$ integers, then we have a discovery; otherwise, not. 
  \item Repeat Steps~1 to~3 $N_\mathrm{obs}$ times, and count the number of discoveries, $N_\mathrm{disc}$. 
  \item Compute the empirical probability, $N_\mathrm{disc}/N_\mathrm{obs}$, which was 0.156845, 0.157050, and 0.156695 in three runs we performed using the parameters above. These values compare well with the theoretical value, ${\cal P}_{2,j}$ and confirm Eq.~12. (As a benchmark, one run required 20,000 sec on a 3~GHz Intel Xenon processor running \textsc{Mathematica}~6.) 
\end{enumerate}

\section{Applications of \boldmath{${\cal P}$} to Observing Programs}\label{newsec4}

Two applications of ${\cal P}$ and Eq.~(\ref{eq12}) must be sharply distinguished. The first application is in the scheduling algorithm for real or simulated observing programs, where we use the discovery rate, 
\begin{equation}
{\cal Z}_{i,j}=\frac{{\cal P}_{i,j}}{t_\mathrm{exp}+OH}~,
\label{new-eq13}
\end{equation}
as a merit function or science benefit/cost metric for optimizing the observing program for discovery. In this application, $\eta$ in Eq.~(12) is $\eta_\mathrm{ops}$. (OH is any observational overhead time that will be charged to the program, such as for calibration or alignment.) 

The second application of Eq.~(\ref{eq12}) is in estimating the probability distribution of discoveries for simulated observing programs, as discussed below. In this application, $\eta =\eta_\mathrm{true}$, where $\eta_\mathrm{true}$ is the ``true'' value. In this context, $\eta_\mathrm{true}$ is a control parameter. If $\eta_\mathrm{ops}\ne\eta_\mathrm{true}$, the number of planets discovered by an observing program may be less than optimal, because the scheduling algorithm may not always choose, as the next star to observe, the qualified star with the highest ``true'' value of the merit function.

In both applications of Eq.~(\ref{eq12}), $C_{i-1,j}$ is the accumulated completeness from all LSOs of star $j$ prior to the $i^{\mathrm{th}}$, each contribution computed as accurately as desired from Eq.~(\ref{eq5}).

\subsection{Probability Distribution of the Number of Discoveries, \emph{pdf}(\emph{m})}\label{sec6}

No matter what star $j$, nor what search $i$ of that star, the $k^{\mathrm{th}}$ LSO in the overall observing program discovers $u$ planetary systems, where $u \in \{0,1\}$ is a Bernoulli random variable with probability ${\cal P}_{k}$, as given in Eq.~(\ref{eq12}) using $\eta_\mathrm{true}$ and the indices $i$ and $j$ corresponding to the $k^{\mathrm{th}}$ LSO.  [For each star ($j$), multiple visits ($i$) are possible, so both $i$ and $j$ define the $k^{\mathrm{th}}$ LSO.]  The probability density function 
(\textit{pdf}) of $u$ for the $k^{\mathrm{th}}$ LSO is 
\begin{equation}
\label{eq13}
\textit{pdf}_k (u)=(1-{\cal P}_k )\delta (u,0)+{\cal P}_k \delta (u,1)~,
\end{equation}
where $\delta(i,j)\equiv 1$ for $i=j$, and zero otherwise, is the Kronecker delta.

An entire observing program, consisting of $n_\mathrm{total}$ LSOs, where
\begin{equation}
\label{eq14}
n_\mathrm{total} =\sum\limits_{j=1}^{n_\mathrm{stars} } {n_j }~,
\end{equation}
where $n_j$ is the total number of LSOs of star $j$, and where $n_\mathrm{stars}$ is the total number of stars observed, discovers $m$ planets, where $m\in\{0, 1, {\ldots} n_\mathrm{total}\}$ is the sum of $n_\mathrm{total}$ Bernoulli random variables, each with $\textit{pdf}_{k}$ given by Eq.~(\ref{eq13}). Therefore, the \textit{pdf} of $m$ is the convolution ($\star$) of $\textit{pdf}_{k}$ for all $k$: 
\begin{equation}
\label{eq15}
\textit{pdf}\,(m)=\textit{pdf}_1 (u)\star \textit{pdf}_2 (u)\star\ldots\star \textit{pdf}_{n_\mathrm{total} } (u)~,
\end{equation}
where each successive convolution has the form
\begin{equation}
\label{eq16}
(\textit{pdf}_k \star{\cal F})(n)\equiv (1-{\cal P}_k ){\cal F}(n)+{\cal P}_k {\cal F}(n-1)~.
\end{equation}

Equation~(16) offers a practical advantage for estimating the outcome of a search program. Starting from a \textit{single} simulated observing program, it allows us to estimate theoretically the \textit{pdf} of the total number of discoveries. The alternative---running many full simulations to build up an empirical estimate of the \textit{pdf} from the discovery results---is much less efficient.

\subsection{Estimation of \boldmath{$\eta$}}\label{sec7}

The pertinent record of a real or simulated observing program is a set of 
$n_\mathrm{total}$ data triplets re-indexed from $i$ (LSO) and $j$ (star) to $k$ (observation):
\begin{equation}
\label{eq17}
\left\{ {c_k ,\ C_{k},\ u_k } \right\}
\equiv \left\{ {c_{i,j},\ C_{i-1,j},\ u_{i,j} } \right\}~,
\end{equation}
for $1\le k \le n_\mathrm{total}$, where $u_{i,j}=0$ or 1 is the number of discoveries by the $i^{\mathrm{th}}$ LSO of the $j^{\mathrm{th}}$ star. (If we assume that we stop searching after a discovery, $u_{i,j}=1$ for at most one value of $i$ for any $j$.)

The logarithmic likelihood function ${\cal L}$ is the logarithm of the probability of the set $\{u_{k}\} \equiv \{u_{i,j}\}$ as a function of $\eta$:
\begin{equation}
\label{eq18}
{\cal L}(\{u_k\}\vert \eta )=
\sum\limits_{k=1}^{n_\mathrm{total} } {\ln p_{u_k } } =
\sum\limits_{k=1}^{n_\mathrm{total} } {\ln \left( {\frac{c_k \eta }{1-
C_{k}\eta }u_k +\left( {1-\frac{c_k \eta }{1-C_{k}\eta }} \right)
(1-u_k )} \right)} ~.
\end{equation}
The maximum-likelihood estimate of the occurrence probability, $E(\eta)$, is the $\eta$-root of the equation:
\begin{equation}
\label{eq19}
\frac{\partial {\cal L}(\{u_k \}\vert \eta )}{\partial \eta }=0~.
\end{equation}
The minimum variance bound (MVB) is the inverse of the Fisher information 
near $E(\eta)$:
\begin{equation}
\label{eq20}
\mathrm{MVB}\left(E(\eta)\right)=\left( {\left. {-\frac{\partial ^2{\cal 
L}(\{u_k \}\vert \eta )}{\partial \eta^2}} \right|_{\eta = E(\eta)} } 
\right)^{-1}~.
\end{equation}

We want to confirm that Eqs. (19--21) accurately estimate $\eta$ and its variance.  To that end, we performed a numerical experiment simulating 100,000 missions of 100 LSOs, according to the following steps.
\begin{enumerate}
  \item Generated $n_\mathrm{total}=100$ random data triplets for the left side of Eq.~(\ref{eq17}) as follows:
\begin{eqnarray}
c&=&{\cal R}~,\label{eq21}\\
\noalign{\smallskip}
C&=&(1-c){\cal R}~,\label{eq22}\\
u&=&{\cal B}\left( {\frac
{\eta c}{1-\eta C} 
}\right)~,\label{eq23}
\end{eqnarray}
where ${\cal R}$ is a uniform random deviate on the range 0--1, ${\cal B}(p)$ is a Bernoulli random deviate with probability $p$, and $\eta = 0.10$. 

\item Compute $E(\eta)$ and MVB($E(\eta)$) using Eqs.~(19--21). 
\end{enumerate}

For the sample of 100,000 trials, we found
\begin{equation}
\langle E(\eta)\rangle=0.101~, 
\label{eq25}
\end{equation}
\begin{equation}
\sigma_{\eta} =0.039~, 
\label{eq26}
\end{equation}
\indent and
\begin{equation}
\left\langle\sqrt{{\rm MVB}(E(\eta))}\right\rangle=0.042~.
\label{eq27}
\end{equation}

Equation~(\ref{eq25}) is the mean value of $\eta$ found using Eq.~(\ref{eq19}) in the 100,000 simulated missions. Equation~(\ref{eq26}), the standard deviation of those values of $\eta$, is the empirical estimate of the scatter in $\eta$ determined by Eq.~(\ref{eq19}). The square of Eq.~(\ref{eq26}) is the empirical variance. The Cram\'er-Rao theoretical limit on the variance of any estimator of $\eta$ is the MVB given by Eq.~(\ref{eq20}). We computed the MVB for each simulated mission, and Eq.~(\ref{eq27}) gives the mean value of the square root of the MVB for the suite of 100,000 simulated missions---computed for direct comparison with the empirical value in Eq.~(\ref{eq26}). 

These results illustrate that the maximum likelihood estimator accurately recovers $\eta$ from a record of the results of observing programs, and that the accuracy of this estimator appears to approach the Cram\'{e}r-Rao limit ($\sigma^2\approx$~MVB). (This Monte Carlo experiment required 270~sec on 56 2.66~GHz Intel Xenon processors operating in parallel.)

\section{Illustrative Design Reference Missions (DRMs)}\label{sec8}

The purpose of a design reference mission (DRM) is to gauge the science 
operations of a mission concept. To illustrate the new completeness methods 
introduced in this paper, we now describe a ministudy using simple DRMs to explore and measure of the power of the \textit{James Webb Space Telescope} (\textit{JWST}) to discover and characterize Earth-like extrasolar planets using a starshade to suppress scattered starlight \citep{Cash,Soummer}. In this scenario, \textit{JWST }and the starshade revolve in coordinated orbits around the second Earth-Sun Lagrange point, L2. The starshade operates on a $\sim$70,000~km sphere centered on \textit{JWST}. In a 3-year planet-finding mission, we assume enough propulsion to slew the starshade 70~times to take up new positions between \textit{JWST} and target stars. We want the DRMs to tell us about the science, for example how many discoveries to expect if we optimize the observing program, assuming $\eta_\mathrm{true}=\eta_\mathrm{ops}=0.3$, say.

Other DRM inputs include: a science strategy; a definition of POIs (same as 
Section~\ref{sec3}, with $q$ depending on filter as given in Table~1); \textit{IWA}~= 0.085~arcsec, $\Delta\textit{mag}_{0} = 26$, and pointing restrictions $\gamma _{1}=85^{\circ}$ (solar avoidance) and $\gamma_2=105^{\circ}$ (starshade bright-side avoidance); an input catalog of stars; exposure time calculators; typical overheads \textit{OH}; and a merit function---in this case, the discovery rate ${\cal Z}$---which we use to select the next star to search.

The science strategy is simple: we perform an LSO, and if a likely POI is 
discovered, we immediately perform additional spectrophotometry to 
characterize the body. Such immediate follow-up reduces the risk of a newly 
discovered POI becoming undetectable before it can be characterized, and 
avoids the difficulty of trying to recover it with inadequate knowledge of 
its orbit \citep{BSH07}. If we discover a POI, we cease further LSOs of that star. If we do not find one, we move the starshade to the next target star, but we return to a star already searched if and when it once again offers the highest value of ${\cal Z}$.

The LSO is a deep image using whichever NIRCam filter offers maximum ${\cal Z}$.  We call this filter the preferred filter, and it varies from star to star. The possible filters are listed in Table~1.

After an LSO finds a potential POI, we obtain images through the four 
\textit{non-preferred} NIRCam filters for that star, and take a low-resolution spectrum with NIRSpec. We use exposure time calculators for NIRCam and NIRSpec, based on parameters from the instrument teams, to achieve $S/N=5$ on a source of magnitude $\textit{mag}_{j}+\Delta\textit{mag}_{0}$ for the LSOs, and $\textit{mag}_{j}+\Delta\textit{mag}_\mathrm{median}$ for the follow-up filter photometry and spectroscopy, where $\textit{mag}_{j}$ is the apparent magnitude of star $j$, and $\Delta\textit{mag}_\mathrm{median}$ is the median magnitude difference between the star and the universe of possibly detected POIs for that star. (If $\Delta\textit{mag}_\mathrm{median}>\Delta\textit{mag}_{0}$, we use $\Delta\textit{mag}_{0}$.) 

At this stage of the study we do not try to refine our understanding of the exposure time calculation beyond the current estimates by the instrument teams (Marsha Rieke et~al.\ and Peter Jakobsen et~al., private communication).  We use standard parameters based on instrument requirements. There may be better observing modes for this particular application (e.g., involving detector sub-arrays).

We interpolate standard stellar magnitudes and zero points to the effective 
wavelengths of each NIRCam filter and the NIRSpec prism, starting from the 
\textit{VJHK} magnitudes from NStED (nsted.ipac.caltech.edu/) and the \textit{VJHK} zero points from \citet{Leinert}. We used a near-infrared spectrum of Earth calculated by Sara Seager (private communication) to estimate the effective geometric albedo $q$ of Earth at the wavelengths and resolving powers of each instrument modes. These parameters are listed in Table~1.

As illustrated in Figure~3, each potential target star is continuously 
observable only for a limited period of time, once or twice a year, 
depending on its ecliptic latitude ($b$). The total time costs of observing a 
star---itemized in Table~2---must fit into a single observability period for 
that star. Some 26 of the 117 target stars \citet{Brown05} used in an earlier 
study of the coronagraphic \textit{Terrestrial Planet Finder} (\textit{TPF--C}) are qualified according to this criterion. These stars constitute the input catalog for these DRMs (Table~3).

\begin{figure}
\epsscale{.9}
\plotone{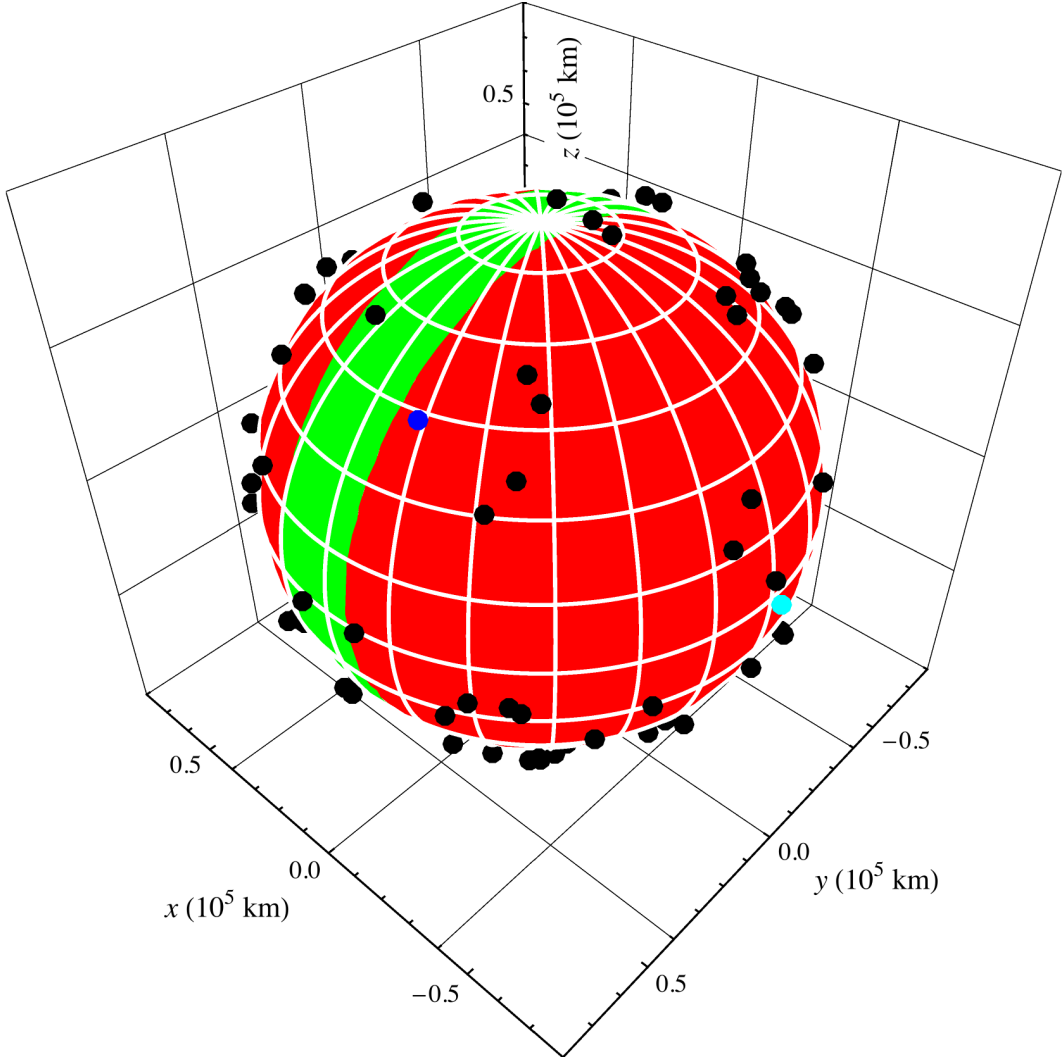}
\label{fig3}
\caption{The sphere of starshade operations, centered on \textit{JWST}, shown here on the vernal equinox. Green: permitted pointings for $\gamma _1=85^{\circ}$ and $\gamma_2=105^{\circ}$. Red: forbidden pointings. As seen from above and as time passes, stars revolve on the starshade sphere around the $+\hat{z}$ axis (north ecliptic pole) in the clockwise direction. Depending on a star's ecliptic latitude $b$, it may be observable for one or two periods per year, or for the entire year ($b>85^{\circ}$). Blue: a typical \textit{TPF-C} target star from \citet[][HIP 92043, $b=+43.4^{\circ}$]{Brown05}. Black: other \textit{TPF-C} stars. Cyan: the Sun, which is fixed in this L2 coordinate system ($-\hat{x}$ toward the Sun, $+\hat{y}$ in the direction of the Earth's orbital motion). }
\end{figure}

To determine the preferred filter for LSOs, we use Eq.~(5) and the 
procedures of Section~\ref{sec3} to calculate virgin completeness $c_{1,j}$ for each of the 26 qualified stars using samples of $N_{0}=40,000$ POIs. We do this separately for each of the five filters, because of the dependence of $q$ on wavelength and resolving power. Next, we use Eqs.~(12--13) to compute ${\cal P}_{1,j}$ and ${\cal Z}_{1,j}$, using \textit{OH}~= 10~hours as our  estimate of the time cost of the fine alignment of the starshade, which we assume is incurred by each new observation with a different instrument, or of a different star. For each star~$j$, we select the filter with the highest value of ${\cal Z}_{1,j}$ as the preferred filter for LSOs of that star (listed in Table~3).

Next, we determine the universe of possible LSOs. For the preferred filter 
only, we continue to compute $c_{i > 1,j}$ using Eq.~(5) until the sample of $N_{0}$ POIs is effectively exhausted. This yields 26 lists of dynamic completenesses, in sequence, one list for each of the 26 stars---some 2,075 values of $c_{i,j}$ in all. Again using Eqs.~(12--13), we convert these lists of $c_{i,j}$ into a full list of possible LSOs in the form of vectors \{HIP$_{k}$, $i_{k}$, ${\cal P}_{ k}$, ${\cal Z}_{k}$\}, where $1\le k\le 2075$ is the index for LSOs introduced in Section~\ref{sec6}, and the items are the Hipparcos number of the star, the number of this visit to that star, and the discovery probability and rate for that visit. 

At the start of a DRM, the prioritized observing program is the list of 2,075 LSO vectors sorted in descending order of ${\cal Z}_{k}$. (Table~4 lists the top 80 LSOs for the current illustration.) We expect to perform only 70 LSOs---but we do not know which ones. How far down the list a DRM reaches is determined by the random discoveries as the DRM unfolds. That is, we determine the outcome of each LSO in turn---discovery, yes or no, with possible follow-on observations and alternative time costs in Table~2---by interrogating a Bernoulli random deviate with probability ${\cal P}$, and interpreting ``1'' as a discovery and ``0'' as no discovery. Each discovery deletes from the observing program all the pending LSOs of a 
star---meaning those with $i$ greater than the visit number~$i$ of the LSO that produced the discovery. These deletions promote the lower priority LSOs of other stars into higher positions on the list. To investigate this behavior and its ramifications, we conduct a Monte Carlo experiment of 500,000 DRMs. 

\begin{figure}
\plotone{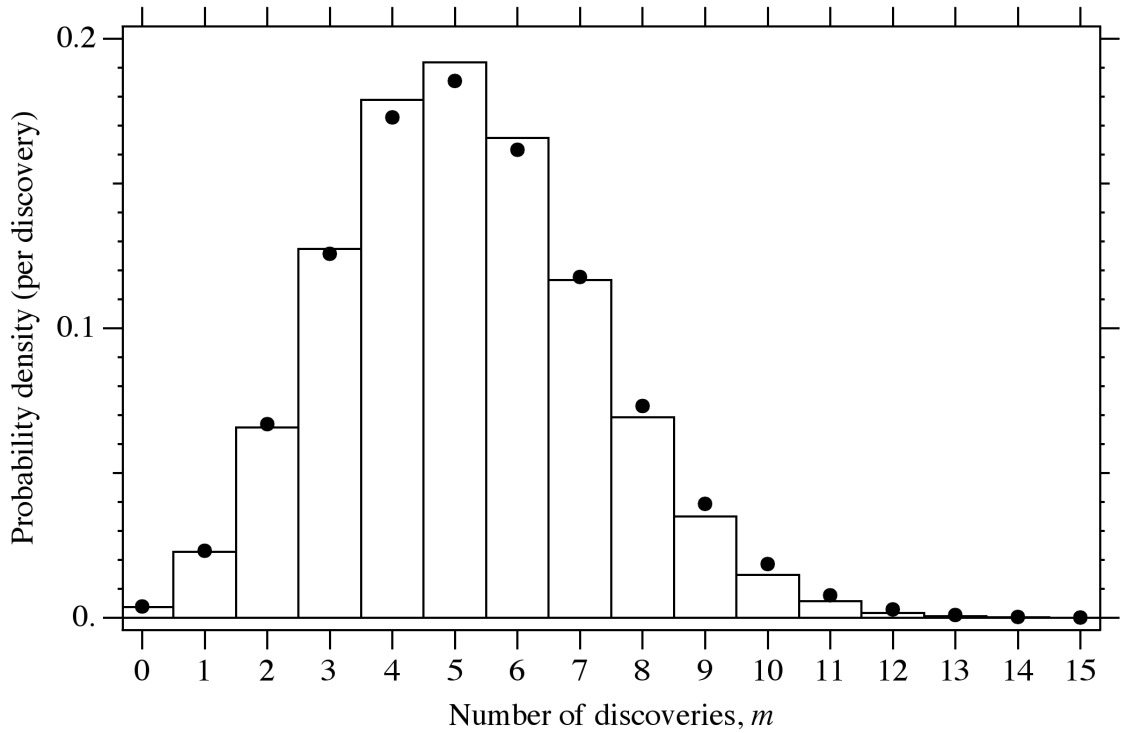}
\label{fig4}
\caption{Comparison of the theoretical and empirical \textit{pdf}($m$). Dots: the theoretical \textit{pdf} estimated from the 70 values of ${\cal P}_{k}$ of the actual LSOs in the single example DRM in Table~4. Histogram: the empirical \textit{pdf} from the actual results for $m$ in 500,000 DRMs. (The histogram has been normalized to 1.0.)}
\end{figure}

Table~4 provides an example DRM in the form of the LSOs actually 
executed in one DRM run, and the discoveries actually made---five in this case. We use it to illustrate the method of deriving the theoretical probability distribution of $m$ from the probabilities ${\cal P}_k$ of a single DRM. We collect the 70 values of ${\cal P}_{k_\mathrm{DRM}}$ for $1\le k_\mathrm{DRM} \le 70$ in Table~4 and follow the recipe in Section~\ref{sec6} to obtain the theoretical \textit{pdf}($m$) represented by the dots in Figure~4. For comparison, the histogram in Figure~4 shows the empirical \textit{pdf}($m$) based on the actual values of $m$ from 500,000 DRMs. 

Table~5 compares the means, standard deviations, and standard deviations of the means---$\langle m \rangle$, $\sigma_{m}$, and 
$\sigma_{\langle m \rangle}$---for the results ensuing from Table~4, which encompass both the example DRM and the 500,000 DRMs computed from the same suite of potential or actual LSOs ($1\leq k\leq 2075$ or $1\leq k_\mathrm{DRM} \leq 70$). The highest precision is achieved by the mean of the 500,000 individual theoretical results for $\langle m\rangle$. We see that our new method produces accurate theoretical estimates of $\langle m\rangle$ and $\sigma_{m}$ from a single DRM and avoids the onerous alternative, which is to conduct a large number of DRMs to obtain empirical results.

The 500,000 DRMs ensuing from the prioritized LSOs in Table~4 offer 
additional insights into the science operations of the \textit{JWST} starshade mission. The most likely number of unique stars searched is 23 (ranges 22--25). The median number of visits per star is 3. The total \textit{JWST} observing time used by a DRM is $7.6\pm1.3\times10^{6}$~sec, and the total number of LSOs is 70 (here, this is the limiting parameter). Note that the total observing time corresponds to only $\sim$7\% of the total \textit{JWST} observing time. The most likely probability of zero discoveries is 0.004 (ranges 0.003 to 0.006).

Table 6 gives the data triplets defined in Eq.~(\ref{eq17}), which we use to  
estimate $\eta_\mathrm{true} = 0.30 \pm 0.12$, where the quoted error is  
the square root of the MVB.

A last note on this ministudy. We have not treated the recovery times of 
$c_{i,j}$ discussed in Section~\ref{sec4} because these example DRMs are ``dilute,'' meaning the LSOs account for only a small fraction of all the observations of \textit{JWST}. In a ``dense'' DRM, with every LSO competing to be the next observation of the telescope, recovery time is important. Here, however, we can assume that a buffer of at least $\sim$10$^{7}$~sec can conveniently pad the time between any two visits of the same star, which ensures adequate $c_{i,j}$ recovery.

\section{Summary}\label{sec9}

In this paper, we have extended completeness-based metrics and algorithms for direct exoplanet searches to include multiple visits, estimating the probability distribution of search results, and estimating the occurrence rate of extrasolar planets. These extensions open the way for improved scheduling decisions, more realistic expectations for candidate instruments and missions, and enhanced science returns. 

Preliminary DRMs for a starshade mission to enable Earth-like planet 
searches with \textit{JWST} suggest a viable program for $<\!10^{7}$~sec of observing time. About five discoveries and spectral characterizations are expected if $\eta=0.3$. Soon, we hope, the \textit{Kepler} mission will provide a first estimate of the true value of $\eta$.

\acknowledgements
The authors thank Jay Anderson, Massimo Robberto, Elizabeth Barker, Marsha Rieke, and Doug Kelly for help with the NIRCam parameters; Jeff Valenti, Jason Tumlinson, Peter Jakobsen, and Pierre Ferruit for help with the NIRSpec parameters; and Webster Cash, Amy Lo, and Tiffany Glassman for help with starshade parameters. They thank Sara Seager for discussions and providing the atmospheric models used to calculate the albedos of Earth. Thanks to Christopher Burrows and Dmitry Savransky for discussions. Appreciation to Marc Postman and Matt Mountain for encouraging this investigation. Gratitude to Christian Lallo for his help keeping R.A.B.'s computers running. Kudos to Sharon Toolan for preparing the manuscript with such great skill.

\begin{deluxetable}{cccccc}
\tablewidth{0pt}
\tablecolumns{6}
\tablecaption{Spectrophotometric parameters for calculating completenesses and exposure times.}
\tablehead{&&\colhead{Nominal $\lambda$} &\colhead{Resolving} &\colhead{Zero point} &\colhead{Earth geometric}\\
\colhead{Instrument} &\colhead{Mode} &\colhead{(nm)} &\colhead{power} 
&\colhead{(Jy)} &\colhead{albedo ($q$)}
}
\startdata
&F070W &\phn700 &\phn4 &3043 &0.232\\
&F115W &1150 &\phn4 &1766 &0.187\\
NIRCam &F140M &1400 &10 &1324 &0.021\\
&F150W &1500 &\phn4 &1188 &0.103\\
&F162M &1625 &10 &1045 &0.179\\
NIRSpec &prism &1150 &31\rlap{.7} &1766 &0.260\\
\enddata
\tablecomments{Zero points interpolated from values for \textit{VJHK} filters (Leinert 1997). We adjusted the Earth's effective albedo for prism 
spectroscopy at 1150 nm.}
\end{deluxetable}

\begin{deluxetable}{cp{3in}cc}
\tablewidth{0pt}
\tablecolumns{4}
\tablecaption{Observing sequence and typical time costs for a single visit of the starshade to a target star. }
\tablehead{\colhead{Step} &\colhead{Activity} &\colhead{Observing time cost} &\colhead{Clock time cost}
}
\startdata
1 &Final alignment of \textit{JWST}, starshade, and target star &10 hours &10 hours\\
2 &LSO through preferred NIRCam filter &$4.8\times10^{4}$ sec &$4.8\times10^{4}$ sec\\
3 &Analyze data for discovery &&7 days\\
4 &Final alignment of \textit{JWST}, starshade, and target star &10 hours &10 hours\\
5 &Images through four non-preferred NIRCam filters &$1.4\times10^{5}$ sec &$1.4\times10^{5}$ sec\\
6 &Analyze data to prepare for spectroscopy &&7 days\\
7 &Final alignment of \textit{JWST}, starshade, and target star &10 hours &10 hours\\
8 &Spectrum using NIRSpec &$1.9\times10^{5}$ sec &$1.9\times10^{5}$  sec\\
&\multicolumn{1}{r}{Total with discovery} &$4.9\times10^{5}$ sec &$1.7\times10^{6}$ sec\\
&\multicolumn{1}{r}{Total without discovery} &$8.4\times10^{4}$ sec &$6.9\times10^{5}$ sec\\
\enddata
\tablecomments{Overhead costs (1, 3, 4, 6, 7) are fixed. 
Exposure times (2, 5, 8) are median values for the 26 stars in the input 
catalog. With no discovery in step 3, only steps 1--3 are executed during a 
visit to a target star. With a discovery, all 8 steps are executed. During 
the analysis steps (3, 6), \textit{JWST} conducts observations for other programs; only the starshade remains aligned with the target star for this program.}
\label{tab2}
\end{deluxetable}

\begin{deluxetable}{crlccrlllllllll}
\rotate
\tablewidth{0pt}
\tablecolumns{14}
\tablecaption{The input catalog of target stars, ranked in descending order of 
the discovery rate on the first LSO (${\cal Z}_{1}$). }
\tablehead{&&&&&&&&&\colhead{Max}\\
\colhead{Disc.} &&&&\colhead{Distance} &&\colhead{LSO} &\colhead{LSO} 
&\colhead{Max} &\colhead{cont.}\\
\colhead{rate} &\colhead{HIP} &\colhead{Type} &\colhead{$L$} 
&\colhead{(pc)} &\colhead{$b(^{\circ})$} &\colhead{filter} 
&\colhead{$t_{\rm exp}$} &\colhead{time} &\colhead{obs.} 
&\colhead{$c_1$} &\colhead{$C_{\infty}$} &\colhead{${\cal P}_1$} &\colhead{${\cal Z}_1$}
}
\startdata
\phn1 &71681 &K1 V &0.6\rlap{1} &1.35 &$-$43 &F070W &3.2 &6.12 &6.38 &0.8 &1 &0.24 &$-$5.19\\
\phn2 &8102 &G8 V &0.4\rlap{7} &3.65 &$-$25 &F115W &4.03 &6.13 &6.29 &0.76 &1 &0.23 &$-$5.31\\
\phn3 &71683 &G2 V &2.2 &1.35 &$-$43 &F070W &2.43 &6.12 &6.38 &0.55 &1 &0.17 &$-$5.34\\
\phn4 &3821 &G0 V &1.2 &5.95 &47 &F115W &3.94 &6.14 &6.41 &0.59 &1 &0.18 &$-$5.4\\
\phn5 &99240 &G6/8 IV &1.5 &6.11 &$-$45 &F115W &4.12 &6.16 &6.4 &0.56 &1 &0.17 &$-$5.47\\
\phn6 &108870 &K4/5 V &0.2 &3.63 &$-$41 &F150W &4.48 &6.13 &6.37 &0.65 &1 &0.19 &$-$5.53\\
\phn7 &22449 &F6 V &2.6 &8.03 &$-$15 &F115W &3.88 &6.15 &6.26 &0.39 &0.96 &0.12 &$-$5.57\\
\phn8 &19849 &K0/1 V &0.4\rlap{1} &5.04 &$-$28 &F115W &4.69 &6.16 &6.3 &0.64 &1 &0.19 &$-$5.64\\
\phn9 &15510 &G8 III &0.7\rlap{1} &6.06 &$-$58 &F115W &4.68 &6.18 &6.53 &0.63 &1 &0.19 &$-$5.65\\
10 &2021 &G1 IV &3.9 &7.47 &$-$65 &F070W &4.2 &6.15 &6.63 &0.34 &0.92 &0.1 &$-$5.7\\
11 &27072 &F6.5 V &2.3 &8.97 &$-$46 &F115W &4.47 &6.25 &6.4 &0.41 &0.98 &0.12 &$-$5.73\\
12 &1599 &G0 V &1.2 &8.59 &$-$58 &F115W &4.71 &6.25 &6.53 &0.5 &1 &0.15 &$-$5.76\\
13 &64394 &G0 &1.3 &9.15 &33 &F115W &4.81 &6.3 &6.32 &0.47 &1 &0.14 &$-$5.85\\
14 &57757 &F8 &3.4 &\llap{1}0.9\phn &0.69 &F115W &4.33 &6.23 &6.24 &0.27 &0.88 &0.08 &$-$5.85\\
15 &12777 &F8 &2.2 &\llap{1}1.2\phn &32 &F115W &4.64 &6.31 &6.32 &0.36 &0.99 &0.11 &$-$5.87\\
16 &96100 &G9 V &0.4\rlap{1} &5.77 &81 &F115W &4.99 &6.2 &7.42 &0.56 &0.97 &0.17 &$-$5.9\\
17 &105858 &F7 V &1.4 &9.22 &$-$47 &F115W &4.91 &6.33 &6.41 &0.47 &1 &0.14 &$-$5.92\\
18 &73184 &K4 V &0.2\rlap{6} &5.91 &$-$4.4 &F150W &5.05 &6.2 &6.25 &0.34 &0.76 &0.1 &$-$6.17 \\
19 &70497 &F8 &3.9 &\llap{1}4.6\phn &60 &F115W &4.75 &6.41 &6.56 &0.17 &0.81 &0.051 &$-$6.26\\
20 &23693 &F6/7 V &1.4 &\llap{1}1.7\phn &$-$79 &F115W &5.18 &6.53 &7.42 &0.33 &0.91 &0.098 &$-$6.29\\
21 &77952 &F0 III/IV &9.6 &\llap{1}2.3\phn &$-$42 &F070W &4.3 &6.23 &6.38 &0.073 &0.4 &0.022 &$-$6.41\\
22 &29271 &G6 V &0.8\rlap{3} &\llap{1}0.2\phn &$-$82 &F115W &5.41 &6.53 &7.43 &0.32 &0.81 &0.095 &$-$6.49\\
23 &114622 &K3 V &0.2\rlap{8} &6.53 &55 &F150W &5.34 &6.27 &6.49 &0.27 &0.68 &0.081 &$-$6.5\\
24 &50954 &F2/3 IV/V &5\phm{.0} &\llap{1}6.2\phn &$-$68 &F115W &4.98 &6.57 &6.7 &0.11 &0.68 &0.032 &$-$6.61\\
25 &40702 &F5 V &6.8 &\llap{1}9.4\phn &$-$75 &F070W &5.29 &6.66 &7.41 &0.076 &0.61 &0.023 &$-$7.01\\
26 &86614 &F5 &5.5 &\llap{2}2\phm{.00} &84 &F070W &5.6 &6.93 &7.46 &0.048 &0.67 &0.014 &$-$7.48\\
\enddata
\label{tab3}
\tablecomments{Times in log seconds. ``Max time'' is the time cost of the full observing sequence in Table~1. ``Max cont.\ obs.'' is the maximum continuous observing time for ecliptic latitude $b$ for $\gamma_{1} = 85^{\circ}$ and $\gamma_{2}=105^{\circ}$. ${\cal P}_{1}$ is the probability of a discovery on the first LSO, from Eq.~(12), and ${\cal Z}_1$ is ${\cal P}_1$ divided by the sum of LSO $t_\mathrm{exp}$ and 10 hours for alignment, from Eq.~(13). The value of ${\cal Z}_1$ is given in log discoveries per sec.}
\end{deluxetable}

\begin{deluxetable}{rcrllcc|crcrllcc}
\tablewidth{0pt}
\tablecolumns{15}
\tabletypesize{\scriptsize}
\tablecaption{Record of a typical DRM.}
\tablehead{
\colhead{$k$} &\colhead{$k_\mathrm{DRM}$} &\colhead{HIP} &\colhead{$i$} &\colhead{${\cal P}_k$} &\colhead{${\cal Z}_k$} 
&\colhead{Discovery} &\hspace{-10pt}$\vert$ &\colhead{$k$} &\colhead{$k_\mathrm{DRM}$} &\colhead{HIP} &\colhead{$i$} &\colhead{${\cal P}_k$} 
&\colhead{${\cal Z}_k$} &\colhead{Discovery}
}
\startdata
1 &\phn1 &71681 &1 &0.24 &$-$5.19 &no &&41 &37 &57757 &3 &0.039 &$-$6.17 &no\\
2 &\phn2 &8102 &1 &0.23 &$-$5.31 &yes &&42 &38 &105858 &2 &0.075 &$-$6.19 &no\\
3 &\phn3 &71683 &1 &0.17 &$-$5.34 &no &&43 &\nodata &71683 &4 &0.022 &$-$6.22 &\nodata\\
4 &\phn4 &3821 &1 &0.18 &$-$5.40 &no &&44 &39 &12777 &3 &0.046 &$-$6.24 &no\\
5 &\phn5 &99240 &1 &0.17 &$-$5.47 &yes &&45 &40 &2021 &4 &0.029 &$-$6.26 &no\\
6 &\phn6 &108870 &1 &0.19 &$-$5.53 &no &&46 &41 &70497 &1 &0.051 &$-$6.26 &no\\
7 &\phn7 &22449 &1 &0.12 &$-$5.57 &no &&47 &42 &23693 &1 &0.098 &$-$6.29 &no\\
8 &\phn8 &19849 &1 &0.19 &$-$5.64 &no &&48 &43 &57757 &4 &0.029 &$-$6.30 &no\\
9 &\phn9 &15510 &1 &0.19 &$-$5.65 &no &&49 &44 &1599 &3 &0.043 &$-$6.31 &no\\
10 &10 &71683 &2 &0.079 &$-$5.66 &yes &&50 &45 &96100 &2 &0.065 &$-$6.31 &no\\
11 &11 &2021 &1 &0.10 &$-$5.70 &no &&51 &46 &108870 &3 &0.032 &$-$6.32 &no\\
12 &12 &27072 &1 &0.12 &$-$5.73 &no &&52 &\nodata &8102 &3 &0.022 &$-$6.32 &\nodata\\
13 &13 &1599 &1 &0.15 &$-$5.76 &no &&53 &47 &27072 &4 &0.030 &$-$6.34 &no\\
14 &14 &3821 &2 &0.077 &$-$5.77 &no &&54 &48 &4394 &3 &0.045 &$-$6.35 &no\\
15 &15 &22449 &2 &0.070 &$-$5.79 &no &&55 &49 &71681 &3 &0.017 &$-$6.35 &no\\
16 &\nodata &99240 &2 &0.079 &$-$5.79 &\nodata &&56 &\nodata &99240 &4 &0.022 &$-$6.35 &\nodata\\
17 &16 &71681 &2 &0.054 &$-$5.84 &no &&57 &50 &22449 &5 &0.019 &$-$6.36 &no\\
18 &17 &64394 &1 &0.14 &$-$5.85 &no &&58 &51 &70497 &2 &0.040 &$-$6.37 &no\\
19 &18 &57757 &1 &0.080 &$-$5.85 &no &&59 &\nodata &3821 &4 &0.019 &$-$6.37 &\nodata\\
20 &19 &12777 &1 &0.11 &$-$5.87 &no &&60 &52 &15510 &3 &0.035 &$-$6.38 &no\\
21 &\nodata &8102 &2 &0.059 &$-$5.90 &\nodata &&61 &53 &12777 &4 &0.033 &$-$6.39 &no\\
22 &20 &96100 &1 &0.17 &$-$5.90 &no &&62 &54 &2021 &5 &0.020 &$-$6.41 &yes\\
23 &21 &2021 &2 &0.063 &$-$5.91 &no &&63 &55 &77952 &1 &0.022 &$-$6.41 &no\\
24 &22 &105858 &1 &0.14 &$-$5.92 &no &&64 &56 &105858 &3 &0.045 &$-$6.42 &no\\
25 &23 &27072 &2 &0.073 &$-$5.96 &no &&65 &57 &19849 &3 &0.032 &$-$6.42 &no\\
26 &\nodata &71683 &3 &0.040 &$-$5.96 &\nodata &&66 &58 &57757 &5 &0.020 &$-$6.45 &no\\
27 &24 &22449 &3 &0.044 &$-$6.00 &no &&67 &59 &70497 &3 &0.031 &$-$6.48 &no\\
28 &25 &108870 &2 &0.065 &$-$6.01 &no &&68 &60 &29271 &1 &0.095 &$-$6.49 &no\\
29 &26 &57757 &2 &0.055 &$-$6.02 &no &&69 &61 &73184 &2 &0.048 &$-$6.49 &no\\
30 &27 &1599 &2 &0.075 &$-$6.06 &no &&70 &62 &22449 &6 &0.014 &$-$6.49 &no\\
31 &28 &12777 &2 &0.069 &$-$6.07 &no &&71 &\nodata &71683 &5 &0.012 &$-$6.50 &\nodata\\
32 &29 &15510 &2 &0.071 &$-$6.07 &no &&72 &63 &23693 &2 &0.060 &$-$6.50 &no\\
33 &30 &3821 &3 &0.038 &$-$6.08 &yes &&73 &64 &114622 &1 &0.081 &$-$6.50 &no\\
34 &\nodata &99240 &3 &0.041 &$-$6.08 &\nodata &&74 &65 &27072 &5 &0.021 &$-$6.50 &no\\
35 &31 &2021 &3 &0.042 &$-$6.09 &no &&75 &66 &1599 &4 &0.026 &$-$6.52 &no\\
36 &32 &19849 &2 &0.068 &$-$6.09 &no &&76 &67 &77952 &2 &0.017 &$-$6.52 &no\\
37 &33 &64394 &2 &0.074 &$-$6.13 &no &&77 &68 &12777 &5 &0.022 &$-$6.56 &no\\
38 &34 &27072 &3 &0.047 &$-$6.15 &no &&78 &69 &57757 &6 &0.016 &$-$6.56 &no\\
39 &35 &22449 &4 &0.030 &$-$6.16 &no &&79 &70 &64394 &4 &0.028 &$-$6.56 &no\\
40 &36 &73184 &1 &0.10 &$-$6.17 &no &&80 &\nodata &2021 &6 &0.014 &$-$6.57 &\nodata\\
\enddata
\label{tab4}
\tablecomments{$k$: order of top 80 LSOs in terms of discovery rate ${\cal Z}$; $k_\mathrm{DRM}$: order of 70 LSOs executed in this example DRM; $i$: number of this LSO of this star; ${\cal P}_{k}$: probability of discovering a POI in this LSO; ${\cal Z}_{k}$: log discovery rate of this LSO; ellipses indicate that this LSO was not peformed in this DRM due to a previous discovery.}
\end{deluxetable}

\begin{deluxetable}{cccccccc}
\tablewidth{0pt}
\tablecolumns{8}
\tablecaption{DRM results for $\langle m \rangle$, $\sigma_{m}$, and $\sigma_{\langle m \rangle}$ for the expected number of POIs discovered.}
\tablehead{
&\multicolumn{3}{c}{Empirical} &&\multicolumn{3}{c}{Theoretical}\\
\cline{2-4}\cline{6-8}
&\colhead{$\langle m \rangle$} &\colhead{$\sigma_m$}
&\colhead{$\sigma_{\langle m \rangle}$} & 
&\colhead{$\langle m \rangle$} &\colhead{$\sigma_m$} 
&\colhead{$\sigma_{\langle m \rangle}$}
}
\startdata
Example DRM &5\phm{.132} &\nodata &\nodata &&5.21\phn &2.15 &\nodata\\
500,000 DRMs &5.132 &2.06 &0.003 &&5.129 &0.09 &0.0001\\
\enddata
\label{tab5}
\tablecomments{Empirical results are based on counting the number of discoveries in Monte Carlo experiments where discoveries were decided by Bernoulli random deviates. Theoretical results use the methods of Section~\ref{sec6} to estimate \textit{pdf}($m$) from the individual LSO probabilities (${\cal P}_{k})$, and the results are---or are derived from---that distribution. (The highest precision is achieved by the mean of the 500,000 individual theoretical results for $\langle m\rangle$.) }
\end{deluxetable}

\begin{deluxetable}{rllcc|cllcc|cllc}
\tablewidth{0pt}
\tablecolumns{14}
\tabletypesize{\small}
\tablecaption{Data triplets for estimation of $\eta$ from results of the typical DRM.}
\tablehead{
\colhead{$k$} &\colhead{$c_k$} &\colhead{$C_k$} &\colhead{$u_k$} &&\colhead{$k$} &\colhead{$c_k$} &\colhead{$C_k$} &\colhead{$u_k$}&&\colhead{$k$} &\colhead{$c_k$} &\colhead{$C_k$} &\colhead{$u_k$}
}
\startdata
1 &0.8038 &0 &0 &&25 &0.1759 &0.6484 &0 &&48 &0.1203 &0.6833 &0\\
2 &0.7574 &0 &1 &&26 &0.1681 &0.2672 &0 &&49 &0.04000 &0.9398 &0\\
3 &0.5537 &0 &0 &&27 &0.2123 &0.5008 &0 &&50 &0.04858 &0.7929 &0\\
4 &0.5950 &0 &0 &&28 &0.2043 &0.3599 &0 &&51 &0.1264 &0.1697 &0\\
5 &0.5560 &0 &1 &&29 &0.1932 &0.6266 &0 &&52 &0.08750 &0.8198 &0\\
6 &0.6484 &0 &0 &&30 &0.09530 &0.8049 &1 &&53 &0.08603 &0.6926 &0\\
7 &0.3886 &0 &0 &&31 &0.1171 &0.5334 &0 &&54 &0.05295 &0.7275 &1\\
8 &0.6446 &0 &0 &&32 &0.1827 &0.6446 &0 &&55 &0.07265 &0 &0\\
9 &0.6266 &0 &0 &&33 &0.2121 &0.4711 &0 &&56 &0.1190 &0.6811 &0\\
10 &0.2199 &0.5537 &1 &&34 &0.1264 &0.6184 &0 &&57 &0.07973 &0.8273 &0\\
11 &0.3442 &0 &0 &&35 &0.07820 &0.7147 &0 &&58 &0.05505 &0.6267 &0\\
12 &0.4061 &0 &0 &&36 &0.3361 &0 &0 &&59 &0.09337 &0.2960 &0\\
13 &0.5008 &0 &0 &&37 &0.1120 &0.4353 &0 &&60 &0.3181 &0 &0\\
14 &0.2099 &0.5950 &0 &&38 &0.2154 &0.4657 &0 &&61 &0.1448 &0.3361 &0\\	
15 &0.2067 &0.3886 &0 &&39 &0.1284 &0.5642 &0 &&62 &0.03500 &0.8414 &0\\	
16 &0.1360 &0.8038 &0 &&40 &0.07705 &0.6505 &0 &&63 &0.1806 &0.3261 &0\\
17 &0.4711 &0 &0 &&41 &0.1697 &0 &0 &&64 &0.2701 &0 &0\\
18 &0.2672 &0 &0 &&42 &0.3261 &0 &0 &&65 &0.05180 &0.8227 &0\\
19 &0.3599 &0 &0 &&43 &0.07945 &0.5473 &0 &&66 &0.06625 &0.8257 &0\\
20 &0.5590 &0 &0 &&44 &0.1126 &0.7131 &0 &&67 &0.05485 &0.07265 &0\\
21 &0.1893 &0.3442 &0 &&45 &0.1813 &0.5590 &0 &&68 &0.05650 &0.7786 &0\\
22 &0.4657 &0 &0 &&46 &0.07942 &0.8242 &0 &&69 &0.04180 &0.6818 &0\\	
23 &0.2123 &0.4061 &0 &&47 &0.07785 &0.7449 &0 &&70 &0.06988 &0.8035 &0\\
24 &0.1193 &0.5953 &0 && & & && & &\\
\enddata
\tablecomments{$k$ here is $k_\mathrm{DRM}$ in Table~4. ${\cal P} _{k}$ in Table~4 can be recovered from $c_k$ and $C_k$ here using Eqs.~(12 \& 18) and $\eta = \eta_\mathrm{ops} = 0.3.$}
\label{tab6}
\end{deluxetable}

\end{document}